\documentstyle[aps,preprint,epsfig]{revtex}

\tightenlines

\begin{document}

\title{Quasideuteron states with deformed core} 
\author{A.~F.~Lisetskiy$\,^1$, A.~Gelberg$\,^1$, R.~V.~Jolos$\,^{1,2}$,
 N.~Pietralla$\,^{1,3}$, P.~von Brentano$\,^1$\\}

\address{$^1\,$ Institut f\"ur Kernphysik, Universit\"at zu K\"oln, 
                50937 K\"oln, Germany \\
         $^2\,$ Bogoliubov Theoretical Laboratory, 
         Joint Institute for Nuclear Research, 141980 Dubna, Russia  \\
         $^3\,$ Wright Nuclear Structure Laboratory, Yale University,
         New Haven, Connecticut 06520-8124, USA \\ } 
\date{\today}

\maketitle

\begin{abstract}
  The M1 transitions between low-lying T=1 and T=0 states in 
  deformed odd-odd N=Z nuclei are analyzed in the frames of 
  the rotor-plus-particle model. Using the representation of 
  an explicit coupling of angular momenta we show that strong coupling
  of the quasideuteron configurations to the axially deformed core  
  results in a distribution of the total $0^+ \rightarrow 1^+$ 
  strength among a few low-lying $1^+$ states. Simple analytical
  formulae for B(M1) values are derived. The realization of 
  the M1 sum rule for the low-lying $1^+,T=0$ states is indicated. 
  The calculated B(M1) values are found to be in good agreement 
  with experimental data and reveal specific features of collectivity 
  in odd-odd N=Z nuclei.  
   
\end{abstract}



\narrowtext


 Due to the charge independence of the nuclear force, 
the isovector (T=1) neutron-proton (n-p) and like-nucleon 
(n-n and p-p) interactions have to be indistinguishable, as 
proved by the existence of isospin  multiplets of nuclei.
 However a p-n pair can also exist in a T=0 state for 
which there is no need for equality with the T=1 n-p, p-p and n-n forces. 
 
To understand the nuclear interaction in the T=0 channel considerable effort
has 
recently been made but there are still substantial difficulties in  
understanding the T=0 n-p correlations \cite{Vogel,Wyss1,Mac2,Faes99,Ots98}. 
However many features of the nuclear 
structure in the T=0 channel can be studied without reference to the 
peculiarity of the T=0 interaction.
From this perspective the odd-odd N=Z nuclei, which are unique systems 
with both T=1 and T=0 modes coexisting at low energies, are the best 
laboratory to study the isospin antisymmetric (T=0) states and the transitions
between T=1 and T=0 modes.

The remarkable fact that the magnetic dipole moments of the free 
proton ($\mu_\pi=+2.79\mu_N$) and the free neutron ($\mu_\nu=-1.91\mu_N$) are 
comparable in magnitude and of opposite sign implies that the
magnetic dipole (M1) transition operator is the most appropriate tool for 
the investigation of the transitions from T=0 to T=1 states. Moreover the M1 
operator is very sensitive to the relative orientation 
of single nucleon spin ($s$) and orbital angular momentum ($l$) as is well 
illustrated by the familiar Schmidt values \cite{Schm} for 
magnetic dipole moments of odd-mass nuclei.  The observables which are 
most sensitive to the relative orientation of spin and orbital angular 
momentum are B(M1) values for isovector transitions between quasideuteron 
states in odd-odd N=Z nuclei \cite{Lis99}.
The quasideuteron states were defined in \cite{Lis99} as 
one-proton-one-neutron $[\pi j\times \nu j]^{J,T}_M$ configurations in a 
single 
$j$ orbital coupled to an inert even-even N=Z core nucleus 
in its ground state $J^\pi = 0^+,T=0$. 
The $[\pi j\times \nu j]^{J,T}_M$ multiplet of states 
splits into two sequences with different isospin symmetry: T=0, (J - odd) and 
T=1,( J - even). 
The states with isospin quantum number T=0 and T=1 and spins $J$ and $J+1$ are 
connected by M1 transitions. The B(M1) values for these transitions are 
given by simple analytical formulae \cite{Lis99}:
\begin{eqnarray}
\label{1}
 B(M1;J \rightarrow J+1)=
{3 \over 4 \pi}{J+1 \over 2J+1}\left(\rule{0mm}{5mm}j+1+{J \over 2} \right)\left(\rule{0mm}{5mm} j-{J \over 2} \right) \left(\rule{0mm}{5mm} g_{IV}^j \right)^2 \mu_N^2,
\end{eqnarray}
\begin{eqnarray}
\label{gfactor+}
\mbox{where} \hspace{0.3cm} g_{IV}^j=g_p^j-g_n^j={l+\alpha_q4.706 \over j} 
              \hspace{0.5cm} \mbox{for} 
\hspace{0.5cm} j=l+1/2
\end{eqnarray} 
\begin{eqnarray}
\label{gfactor-}
\mbox{and}\hspace{0.3cm} g_{IV}^j={l+1-\alpha_q4.706 \over j+1} 
          \hspace{0.7cm} \mbox{for} 
\hspace{0.5cm} j=l-1/2,
\end{eqnarray}
where the  values of the orbital $g$-factors are taken to be bare 
($g_p^{l,{\rm bare}}=1$,$g_n^{l,{\rm bare}}=0$) and $\alpha_q$ is a 
quenching factor for the spin bare $g$-factors 
($g_p^{s,{\rm bare}} = 5.58$ and $g_n^{s,{\rm bare}} = - 3.82$).
 The 
positive interference of large spin and orbital parts of the isovector 
$\Delta$T=1 M1 reduced matrix elements in the $j=l+1/2$ case was shown to 
cause a strong enhancement of $\Delta$T=1 M1 transitions between 
quasideuteron states with $j=l+1/2$ in  odd-odd 
N=Z nuclei.  Actually the strongest known M1 
$0^+ \rightarrow 1^+$ transitions between low-lying nuclear states are 
observed in odd-odd N=Z nuclei in the lower part of p-, sd- and 
pf-shells
where the $j=l+1/2$ orbitals play a dominant role. On the contrary, for the 
odd-odd N=Z nuclei in the upper part of the p- and sd-shells, the M1 
$0^+ \rightarrow 1^+$ transitions are strongly suppressed due to the 
destructive interference of spin and orbital parts of the M1 matrix elements 
for one-proton-one-neutron configurations in a single $j=l-1/2$ orbital. 

Analyzing the strong M1, $0^+_1 \rightarrow 1^+_i$ transitions in odd-odd N=Z 
nuclei we noted that in  $^{6}$Li,$^{18}$F and 
$^{42}$Sc the total M1 transition strength from the yrast $0^+,T=1$ state to 
the $1^+,T=0$ states is concentrated  in the $0^+_1 \rightarrow 1^+_1$ 
transition.  This suggests that the structure of 
the low-lying states in these nuclei is dominated by simple quasideuteron 
configurations with $j=l+1/2$. In deformed nuclei, e.g., $^{10}$B, $^{14}$N, 
$^{22}$Na and $^{26}$Al the total M1 strength is fragmented among two or three 
low-lying states  indicating a more complex structure.  

The aim of the present paper is to extend the investigation of the  spherical 
shell model quasideuteron configurations to an axially deformed mean field 
and to explore the fragmentation mechanism of the isovector M1 strength in 
odd-odd N=Z nuclei.  
We will show that the low-energy structure of the deformed odd-odd N=Z nuclei 
where strong M1 transitions are observed could be reasonably well understood 
in terms of a rotor-plus-quasideuteron model and that the electromagnetic 
properties of the low-lying states are strongly affected by the deformed mean 
field. 


The basic assumption of the model \cite{Davidson,Malik} which we apply in this 
paper to the deformed odd-odd N=Z nuclei is that 
one has one proton and one neutron outside of an even-even deformed rotating 
core. We consider the simplified version of this model neglecting the Coriolis 
interaction and 
the residual interaction between the odd proton and odd neutron. 
Then the 
rotational motion of the nucleus is specified by the quantum 
numbers $JMK$ and the total wave function has the form appropriate to a 
rotationally invariant system with axial symmetry which also posses the 
signature symmetry \cite{Bohr}:
\begin{eqnarray}
\label{D}
|JMKT \rangle = \sqrt{2J+1 \over 16 \pi^2 (1+\delta_{K,0})}
\left[\rule{0mm}{5mm} D^J_{MK}\Phi_{K,T}+(-1)^{J+K} D^J_{M-K}\Phi_{\overline{K},T} \right],  
\end{eqnarray}
where $\Phi_{K,T}$ is the wave function in the intrinsic system: 
\begin{eqnarray}
\Phi_{K,T}= { 1 \over \sqrt{2(1+\delta_{1,2})}}\left[\rule{0mm}{5mm} u_\pi(1)u_\nu(2) +(-1)^Tu_\pi(2)u_\nu(1) \right]\cdot\zeta^T_{T_z=0}(1,2), 
\end{eqnarray}
 where $u_\rho(i)$ are single particle eigenfunctions of the Nilsson 
Hamiltonian  with $\Omega_i$ the 3-projection of particle angular momentum, 
$K=\Omega_1+\Omega_2$, $\zeta^T_{T_z=0}(1,2)$ - isospin wave function with  
the isospin quantum number T.
The states belonging to a $K=0$ band have only even (odd) spins for the 
signature quantum number $r=+1$ ($-1$). It can be shown that states belonging 
to the $T=1$ ($T=0$) band have $r=+1$ ($-1$).    
The coupling of the angular momenta of the odd proton, the odd neutron and the
rotor that is implicit in  Eq.(\ref{D})
 can be exhibited by 
transforming to the representation of explicit coupling of 
angular momenta \cite{Bohr} appropriate for a 
strongly coupled system. This representation, which can be treated also as 
an algebraic representation \cite{Guise98}, allows one to work with spherical 
shell model configurations and to study the interplay between different 
possible degrees of freedom generating M1 transitions.     

As a starting point we use particle plus-rotor-model basis states 
written in terms of spherical single-particle wave functions in a 
strong coupling approximation \cite{Bohr,Guise98}:
\begin{eqnarray}
\label{rpat1}
|JMK \rangle  =  
 \sum_{R,j}\sqrt{(1+\delta_{K_R,0})(2R+1) \over 2J+1} C_{RK_Rj\Omega}^{JK}\chi_j^\Omega 
  \left[\rule{0mm}{5mm} 
|R \rangle \otimes |j \rangle \right]^{J}_{M},   
\end{eqnarray}
where $\chi_j^\Omega$ are projection coefficients of single particle 
Nilsson $[Nn_z\Lambda]\Omega$ orbitals on the spherical single particle 
$|nlj\Omega \rangle$ basis \cite{Irvine}:
\begin{eqnarray}
\label{proj}
|Nn_z\Lambda;\Omega\rangle = \sum_{j=\Omega}^{N+1/2}\chi_j^\Omega|nlj\Omega \rangle, 
\end{eqnarray}
$C_{RK_Rj\Omega}^{JK}$ are 
Clebsch-Gordan 
coefficients, R is the core angular momentum quantum number and 
$\left[\rule{0mm}{4mm} |R \rangle \otimes |j \rangle \right]^{J}_{M}=
\sum_{M_R,m}C_{RM_Rjm}^{JM}|R M_R \rangle \cdot |nljm\rangle$
are weakly coupled (SU(2) coupling) rotor-plus-particle states. The wave 
functions for two particle states coupled to the $K_R=0$,T=0 
rotational 
core can be easily constructed applying Eq.(\ref{rpat1}). After some 
simple transformations one obtains:
\begin{eqnarray} 
\label{rpat2} 
|JMKT \rangle  =  
 \sum_{R,J_q}\sqrt{2(2R+1) \over 2J+1} C_{R0J_qK}^{JK}
  \left[\rule{0mm}{5mm} |R \rangle \otimes |J_q \rangle \right]^{JT}_{MT_z=0},
\end{eqnarray}
where the $|J_q \rangle$ is a one-proton-one-neutron state in the deformed 
field: 
\begin{eqnarray}
\label{rpat3} 
|J_q \rangle \equiv |J_q M_qKT \rangle = 
\sum_{j_1,j_2}\chi_{j_1}^{\Omega_1}\chi_{j_2}^{\Omega_2}C_{j_1 \Omega_1 j_2 \Omega_2}^{J_qK}
 \left[\rule{0mm}{5mm} |j_1 \rangle \otimes |j_2 \rangle 
 \right]^{J_qT}_{M_q T_z=0},
\end{eqnarray}
and  $\Omega_i$ is the Nilsson quantum number 
of angular momentum projection on the symmetry axis for the odd proton and 
the odd neutron.

To calculate M1 matrix elements we start with a nuclear magnetic dipole 
operator, which is the sum of proton and neutron one-body terms for spin 
and orbital contributions:
\begin{equation} 
\label{eq:TM1} 
{\bf T}(M1)= \sqrt{3 \over 4 \pi } 
     \left( \sum_{i=1}^{Z}\left[g_p^l{\bf l}^p_i+g_p^s{\bf s}^p_i \right]+
            \sum_{i=1}^{N}\left[g_n^l{\bf l}^n_i+g_n^s{\bf s}^n_i\right] 
     \right)\ \mu_N, 
\end{equation}
where $g_{\rho}^l$ and $g_{\rho}^s$  are the orbital and spin $g$-factors 
and ${\bf l}^\rho_i$, ${\bf s}^\rho_i$ are the single particle orbital angular 
momentum operators and spin operators. 
For the purposes of our paper it is convenient to rewrite the expression
for the M1 transition operator in another form: 
\begin{equation} 
\label{eq:TMR} 
{\bf T}(M1)= {\bf T}_R(M1)+{\bf T}_q(M1),
\end{equation}
where the ${\bf T}_R(M1)$ is an M1 operator for the even-even N=Z core and 
${\bf T}_q(M1)$ is the M1 operator for the odd proton and odd neutron 
subsystem.  
Since only the $T=0, K=0$ states of the even-even N=Z core nucleus are 
assumed to be taken into account, and the angular momentum quantum number $R$ 
can be only even, the core does not contribute to the $\Delta$T=1 M1 matrix 
elements. Taking this into account we can calculate directly the matrix 
elements of the ${\bf T}_q(M1)$ operator defined in the laboratory coordinate 
frame using the wave functions given by  Eq.(\ref{rpat2}). Using the general 
reduction formula for the reduced matrix elements \cite{Bruss77} 
and performing summation over all possible values of R we get:

\begin{eqnarray}
\langle JKT ||T(M1)|| J'K'T' \rangle = \sqrt{2J+1}C_{JK1\nu}^{J'K'} 
\sum_{J_q,J'_q}(-1)^{J'_q}C_{J_qK1\nu}^{J'_qK'} { 
\langle J_qT ||T_q(M1)|| J'_qT' \rangle \over \sqrt{2J'_q+1}},
\end{eqnarray}
where $\nu=K'-K$. This expression clearly shows that configurations with 
various possible $J_q$ values contribute to the 
total M1 $0^+ \rightarrow 1^+$ m.e. with the weight given by the familiar 
Clebsch-Gordan coefficient $C_{J_qK1\nu}^{J'_qK'}$.     
We consider further a particular case assuming
that the initial state is characterized by  K=0 and that T=1 ($J$ even). 
Then, using Eq.(\ref{rpat3}) and well known 
properties of Clebsch-Gordan coefficients and $6-j$ symbols, we get the 
following formula for the  B(M1) values:
\begin{eqnarray}
\label{m1:general}
 B(M1; J,K=0 \rightarrow J',K'=\Omega_1'+\Omega_2')={ \mu_N^2 \over 4 \pi}
\left[\rule{0mm}{5mm}  C_{J01K'}^{J'K'} \sum_{j_1,j_1'}  
\chi_{j_1}^{\Omega_1} \chi_{j_1'}^{\Omega_1'} 
C_{j_1\Omega_1 j_1'\Omega_1'}^{1K'} {\cal M}_{j_1j_1'} \right]^2,
\end{eqnarray}
where ${\cal M}_{jj'}=\sqrt{j(j+1)(2j+1)}g^j_{IV}$ for $j'=j$ 
i.e., for quasideuteron configurations and 
\begin{eqnarray}
\label{Mjj}
{\cal M}_{jj'}=\sqrt{l(l+1)(2l+1)} 
{\left(\rule{0mm}{3mm} g^{j'}_{IV}- g^{j}_{IV} \right) \over 2}
\end{eqnarray}
for spin-orbit partner orbitals with $j=l \pm 1/2, j'= l \mp 1/2$.
 The difference of proton and neutron $g$-factors 
($g_{IV}^{j}=g_{p}^{j}-g_{n}^{j}$) is given by  Eqs.(\ref{gfactor+}) and 
(\ref{gfactor-}). 
$\Omega$ and $\Omega'$ are 
3-projections of particle angular momentum for the Nilsson 
$[Nn_z\Lambda]\Omega$ and $[Nn'_z\Lambda']\Omega'$ orbitals 
used for the construction of the final and initial states involved in 
the transition.   
The terms with $j=j'$ in  Eq.(\ref{m1:general}) represent the individual 
contributions of two nucleon configurations in the single-j-orbital while 
the terms with $j \ne j'$ are related to the single particle isovector 
spin-flip 
( $j=l \pm 1/2 \rightarrow j'=l \mp 1/2$) mechanism. It is interesting to 
note that the isovector spin-flip part is proportional to the difference of 
the isovector $g$-factors for spin-orbital partner orbitals, i.e. it is itself 
decomposed into two different parts produced by the quasideuteron 
configurations with  $j=l + 1/2$ and $j'= l - 1/2$ orbitals. 



\begin{figure}[bt]
\centerline{\epsfig{file=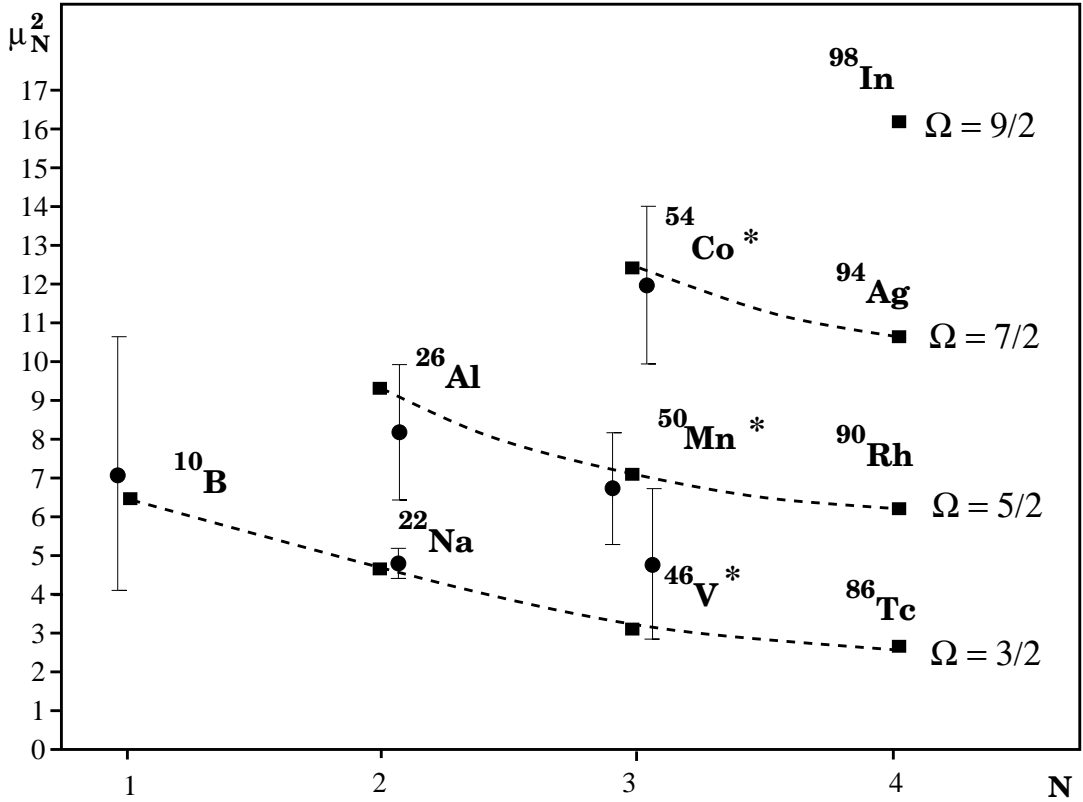,width=0.87\textwidth}}
\vspace{0.2cm}
\caption{Calculated (\protect Eq.~(\ref{bm1:k0}) with $\alpha_q=0.9$) and 
experimental B(M1;$0^+_1,K=0 \rightarrow 1^+_1,K=0$) values as a function 
of principal quantum number N for different values of the $\Omega$ quantum 
number.  The filled circles with error bars and filled 
squares connected with dashed lines are used for experimental and theoretical 
data, respectively.   For the nuclei marked with an asterisk (*), the  
experimental data are deduced following the procedure discussed in the 
text.}
\label{fig1}
\end{figure}

 When the Fermi surface coincides with the Nilsson level with $\Omega \ne 1/2$ 
then it can be stated for a certainty that the 
lowest $1^+$ state in the 
odd-odd N=Z nucleus is a bandhead of a $K=0,T=0$ $(r=-1)$ band 
\footnote{If $\Omega = 1/2$ then the Coriolis interaction mixes 
$\Omega = 1/2$ and $\Omega =-1/2$ single 
particle states and subsequently K=1 and K=0 bands with T=0.}. This means that 
both initial $J$ and final $J'=J+1$ states are characterized by $K'=K=0$
 quantum 
number (i.e. $\Omega'= \Omega$ in Eq.[\ref{m1:general}]) and the expression
(\ref{m1:general}) for the B(M1) values reduces to the following analytical 
form: 
\begin{eqnarray}
\label{bm1:k0}
B(M1; J\rightarrow J+1)= 
{3 \over 4 \pi}\Omega^2\mu_N^2 {J+1 \over 2J+1} \left(\rule{0mm}{5mm} 
\sum_j\chi_j^\Omega\left[\rule{0mm}{7mm}  \chi_j^\Omega \mp 
{\chi_{j \mp 1}^\Omega \over \sqrt{2}}\sqrt{\left(\rule{0mm}{4mm}{l+1/2 \over \Omega}\right)^2-1} \hspace{0.17 cm} \right] g_{IV}^j 
\right)^2,
\end{eqnarray}
where the upper sign is to be used for the $j=l+1/2$ and the lower sign 
for the $j=l-1/2$ case. The property of the ``spin-flip''  m.e. given by  
Eq.(\ref{Mjj}) allows to represent the total M1 m.e. as a sum of partial 
contributions of a single-$j$-orbitals.   
These individual contributions are proportional to the isovector $g_{IV}^j$ 
factor as in the familiar case of two nucleon configurations in a
single-$j$-orbital [ see Eq.(\ref{1})]. 
The $\chi_{j}^\Omega$ coefficients were calculated using the deformation 
parameter $\beta_{\rm eff}$ deduced (see \cite{Bohr}) from known 
B($E2;2^+_1\rightarrow0^+_1$) values (see Table \ref{tab:bm1k0}). 
For the resulting Nilsson wave functions we use the following short notation 
$|[lj],\Omega \rangle$, where  $lj$ indicates the dominant spherical 
orbital in the Eq.~(\ref{proj}).   
Finally, 
B(M1; $0^+_1 \rightarrow 1^+_1$) values for the  deformed odd-odd N=Z nuclei 
with the $1^+_1$ state 
characterized by the K=0 are collected in the Table \ref{tab:bm1k0}.
The experimental B(M1;$0^+_1 \rightarrow 1^+_1$) values for $^{46}$V
\cite{Fries98,Brent01}, $^{50}$Mn \cite{Pietr00} and $^{54}$Co 
\cite{Schneider99} were taken as 
$B(M1;0^+_{T=1,K=0} \rightarrow 1^+_{T=0,K=0})= 
D^{th}B(M1;3^+_{T=0,K=0} \rightarrow 2^+_{T=1,K=0}$) were $D^{th}$ is the 
ratio of the calculated B(M1) values: 
$D^{th}=B^{th}(M1;0^+_{T=1,K=0} \rightarrow 1^+_{T=0,K=0})/B^{th}(M1;3^+_{T=0,K=0}\rightarrow 2^+_{T=1,K=0})$. 
These data are plotted in Figure~\ref{fig1} giving also the 
predictions for the cases where the $g_{9/2}$ orbital (N=4) is expected to be 
dominant. 

First, we conclude from  
 Table \ref{tab:bm1k0} and  Figure~\ref{fig1} a surprisingly good 
agreement of the experimental data with the theoretical results. It indicates 
that to a large extent,  
the structure of the low-lying states in the odd-odd N=Z 
nuclei considered, is determined by the deformed mean field. 

Second, we note from 
Eq.(\ref{bm1:k0}) the proportionality of the 
B(M1) values to $\Omega^2$. This theoretical result is clearly supported 
by the known experimental data as well as by 
full pf-shell model calculations with the KB3 
residual interaction for $^{46}$V \cite{Fries98} and $^{50}$Mn 
\cite{Schmidt}. Following the predictions of the rotor-plus-quasideuteron 
model (see Table \ref{tab:bm1k0}) one finds that the ratio of 
B(M1;$0^+_1\rightarrow 1^+_1)$ values in  $^{46}$V  and $^{50}$Mn is 0.45 
that is very close to the shell model value of 0.44 \cite{Fries98,Schmidt}. 
It shows that B(M1)
values can provide us with additional information on collective states in 
odd-odd N=Z nuclei which can not be obtained from the B(E2) values.    


When one of the two nucleons occupying a Nilsson orbital with 
$\Omega \ne 1/2$ lying on the Fermi surface is moved to the higher orbital 
with $\Omega' =\Omega \pm 1$ or when one of the nucleons from the lower 
lying Nilsson orbital is moved to the orbital on the Fermi surface, then 
one can construct other low-lying $1^+$ states which are the bandheads of 
K=1,T=0 bands. Using Eq.(\ref{m1:general}) for the M1 m.e. we have 
calculated B(M1;$0^+,K=0 \rightarrow 1^+_i,K=1$) values for some of the 
odd-odd N=Z nuclei. We collect our predictions in Table \ref{tab:bm1k0} 
to compare with the known experimental data. While the quality of the 
agreement 
with experiment is good, it is worse than for the K=0 $1^+_1$ states 
discussed above. This indicates larger admixtures of other configurations 
than in the K=0 $1^+_1$ case. From Table \ref{tab:bm1k0} one can see 
that if the Nilsson orbital with $\Omega' =\Omega \pm 1$ contains also 
a large component with $j=l+1/2$ ( similarly to the orbital with 
$\Omega$ quantum number) then one gets a large strength of the $\Delta$K=1, 
$0^+ \rightarrow 1^+_i$ transition. This strength is comparable with the 
strength of the $\Delta$K=0, $0^+ \rightarrow 1^+_i$ 
transition (see, for example, results for $^{22}$Na) discussed above. 
Moreover if one sums the strengths of  $0^+ \rightarrow 1^+$ transitions with 
$\Delta$K=1 and $\Delta$K=0,  one gets the value which is approaching the 
one given by the Eq.~(\ref{1}) for the quasideuteron spherical 
configurations. 
The simplest way to see the realization of a kind of quasideuteron sum 
rule mentioned above is to consider the deformed single j orbital 
approximation. In this 
case the  $\chi_{j}^\Omega$ coefficients with $j\ne N+1/2$ vanish 
while  $\chi_{j=N+1/2}^\Omega$ = 1. Then using Eq.~(\ref{m1:general}) we get 
\begin{eqnarray}
\label{bm1:k01}
B(M1; 0^+_{1,(T=1,K=0)}\rightarrow 1^+_{1,(T=0,K=0)})= 
{3 \over 4 \pi}\Omega^2 \left( g_{IV}^j \right)^2 \mu_N^2, \hspace{0.3cm} 
\mbox{and}
\end{eqnarray} 
\begin{eqnarray}
\label{bm1:k11}
B(M1; 0^+_{1,(T=1,K=0)}\rightarrow 1^+_{(2,3),(T=0,K=0)})= 
{3 \over 8 \pi}\left( j + \Omega_{2,3} \right)
\left( j - \Omega_{2,3}+1 \right)
 \left( g_{IV}^j \right)^2 \mu_N^2,
\end{eqnarray}
where $\Omega_{2,3} =1 \pm |\Omega|$. Summing up the strengths for three 
$1^+$ states we obtain 
\begin{eqnarray}
\label{bm1:k111}
\sum_{i,K}B(M1; 0^+_{1,(T=1,K=0)} \rightarrow 1^+_{i,(T=0,K)})= 
{3 \over 4\pi}j(j+1)\left( g_{IV}^j \right)^2 \mu_N^2,
\end{eqnarray} 
that is exactly the same as the expression yielded by the Eq.~(\ref{1})
for the J=0 case. This exercise shows that one of the consequences of 
deformation is a 
splitting of the quasideuteron states, i.e. the splitting of the single 
particle states and their coupling to the different spins of the deformed 
core result in the appearance of a few low-lying $1^+,T=0$ states 
connected with the lowest $0^+,T=1$ state by comparably strong M1 
transitions.
Another effect caused by the deformation is 
the mixing of the different single $j$ orbitals. This leads to 
the modification of the sum rule discussed above but it does not 
change substantially the whole picture -- the main fragments of the 
quasideuteron strength given by Eq.~(\ref{bm1:k111}) will be still 
concentrated in a few lowest $1^+$ states.  As an 
example we present in Table~\ref{tab:sum} the results of the exact 
calculations using Eq.~(\ref{m1:general}) for M1 m.e. and calculations 
in the deformed single-$j$-orbital approximation using Eqs.~(\ref{bm1:k01}) 
and (\ref{bm1:k11}). 

Very interesting specific case is the one when the odd proton and the odd 
neutron occupy Nilsson orbital with $\Omega=N+1/2$. Then only 
one spherical component, namely with $j=N+1/2$, contributes to the Nilsson 
single particle state (see Eq.~[\ref{proj}]) and subsequently the 
B(M1;$0^+_1 \rightarrow 1^+_1$) value is insensitive to the deformation and 
is given by  Eq.~(\ref{bm1:k01}). Moreover using Eq.~(\ref{m1:general}) one 
can  prove that in this case the sum of B(M1;$0^+_1 \rightarrow 1^+_i$) 
values for all possible $1^+,T=0$ states within single N-oscillator shell 
does not depend on the deformation. This is the case of $^{10}$B, $^{26}$Al 
and $^{54}$Co nuclei. 
The existence of this sum rule for the above discussed specific case but 
for even-even nucleus $^{12}$C was noted recently by L. Zamick and 
N.Auerbach \cite{Zamick2}.

Our present consideration was focused on the M1 $0^+_1 \rightarrow 1^+_i$ 
transitions. However we want to note that the transitions between the states
with spin values different from $J=0$ and  $J=1$ are also of great importance.
Beside the yrast $K=0,T=1$ and $K=0,T=0$ bands the yrast $K=2\Omega,T=0$ band 
is present in odd-odd N=Z nuclei at low energies. Then in the collective model the 
isovector M1 transitions between the states of the $K=0,T=1$ and 
$K=2\Omega,T=0$ (with $\Omega \ge 3/2$) bands are forbidden. Thus these 
forbidden isovector M1 transitions on the background of enhanced isovector 
M1 transitions can be used as effective indicators of the goodness of the $K$ 
as a quantum number in odd-odd N=Z nuclei. For instance, it was illustrated 
recently 
\cite{Erice2} that this collective model selection rule, which arises also in 
large scale shell model calculations, works perfectly in $^{46}$V and 
$^{50}$Mn. 

In the present paper we have avoided discussion of the E2 transition 
strengths which are traditional tools to investigate the collective properties 
of nuclear states. They were frequently used to explore the collectivity 
in odd-odd N=Z nuclei, too. It was well established 
(see, for example, \cite{Malik,Erice2,Wi,Bologna}) that the behavior of 
experimental and shell model E2 m.e.'s in deformed odd-odd N=Z nuclei in
the sd-shell and in the pf-shell is reproduced by the simple geometrical 
model rather well for low-spin states. 

 In conclusion, we have analyzed the properties of the rotor-plus-particle model wave 
functions with respect to the quasideuteron degree of freedom which is 
related to the very strong M1 transitions in odd-odd N=Z nuclei. 
 We have found that strong coupling of the quasideuteron to 
the different states of the deformed core results in several comparably 
strong M1 
$0^+_1 \rightarrow 1^+_i$ transitions. This is in contrast to the simple 
quasideuteron scheme with a $J=0^+$ spherical core where the M1 strength is 
concentrated in one strong M1 transition between the lowest $0^+$ and $1^+$ 
states. The results of calculations are found to be in good agreement with 
experimental data and demonstrate an $\Omega^2$-dependence of B(M1) values 
for $\Delta T=1, \Delta K=0$ transitions.  The existence of sum rule for 
the low-lying $1^+$ states in collective model was demonstrated. 
 The predictions for heavier proton 
rich nuclei indicate that strong enhancement of M1 transitions in odd-odd N=Z
nuclei is to be expected in the exotic region up to $^{100}$Sn.

 
 The authors thank C.~Friessner, A.~Schmidt, I.~Schneider for discussions.
 This work is supported by the DFG under Contracts No. Br 799/10-1 
 and Pi 393/1-1. One of the authors (N.P.) received partial support by the 
 US-DOE under Grant No. DE-FG02-91ER-40609.    

\begin{table}
\caption{The B(M1) values for the transitions between $0^+_1,K=0$ state 
 and $1^+_f,K$ states. The structure of the $1^+_f,K$ states is shown in 
the fourth column, where $[lj]$ indicates the dominant spherical component 
in \protect Eq.(~\ref{proj}). The calculated B(M1) values are 
given for bare spin $g$-factors ($\alpha_q$=1.0) and quenched ones with 
$\alpha_q$=0.9. Experimental B(M1) values shown in the last column are taken 
from \protect\cite{Fries98,Brent01,Pietr00,Schneider99,NS}.}
\label{tab:bm1k0}
\begin{center}
\begin{tabular}{ccccccc}
 Nucleus & $\beta_{\rm eff}$ & State & Structure &
\multicolumn{3}{c}{B(M1;$0^+_1 \rightarrow 1^+_f,K_f$), ($\mu_N^2$)} \\ 
\cline{5-7}
         &                   &  $J^\pi_f,K_f$ &
$|[lj],\Omega \rangle \times |[l'j'],\Omega' \rangle$ &  
\multicolumn{2}{c}{Theory} & Expt. \\
         &         &    &  & $\alpha_q=1.0$ & $\alpha_q=0.9$ &  \\
\cline{3-4} \cline{5-6}
$^{10}$B  & 0.8      & $1^+_1,0$  & 
$|[p_{3/2}],3/2 \rangle \times |[p_{3/2}],3/2 \rangle $ 
                                          &  7.8 & 6.5 &  7.5(32) \\
          &          & $1^+_2,1$  & 
$|[p_{3/2}],3/2 \rangle \times |[p_{1/2}],1/2 \rangle $ 
                                          &  1.7 & 1.5 &  0.59(5) \\
          &          & $1^+_3,1$  & 
$|[p_{3/2}],3/2 \rangle \times |[p_{3/2}],1/2 \rangle $ 
                                          &  6.3 & 5.1 &   - \\ 
$^{22}$Na &  0.43    & $1^+_1,0$  & 
$|[d_{5/2}],3/2 \rangle \times |[d_{5/2}],3/2 \rangle $ 
                                          & 5.4 & 4.6 &  5.0(3) \\
          &          & $1^+_2,1$  & 
$|[d_{5/2}],3/2 \rangle \times |[s_{1/2}],1/2 \rangle $ 
                                          & 5.8 & 4.7 &  4.4(10) \\
          &          & $1^+_3,1$  & 
$|[d_{5/2}],3/2 \rangle \times |[d_{5/2}],1/2 \rangle $ 
                                          & 3.5 & 3.1 &  $>$4.4 \\ 
$^{26}$Al &  0.38    &$1^+_1,0$  & 
$|[d_{5/2}],5/2 \rangle \times |[d_{5/2}],5/2 \rangle $  
                                           &10.7 & 9.3  &  8(2) \\
          &          & $1^+_2,1$  & 
$|[d_{5/2}],5/2 \rangle \times |[d_{5/2}],3/2 \rangle $ 
                                          & 2.8 & 2.5 &  0.8(1) \\ 
$^{46}$V &   0.23    & $1^+_1,0$  & 
$|[f_{7/2}],3/2 \rangle \times |[f_{7/2}],3/2 \rangle $  
                                          & 3.7 & 3.2 &   5(2)\footnote{The 
values represent an experimental estimations (see text).}  \\    
          &          & $1^+_2,1$  & 
$|[f_{7/2}],3/2 \rangle \times |[f_{7/2}],1/2 \rangle $ 
                                          & 7.6 & 6.7 &   \\
          &          & $1^+_3,1$  & 
$|[f_{7/2}],3/2 \rangle \times |[f_{7/2}],5/2 \rangle $ 
                                          & 5.7 & 5.1 &   \\ 
$^{50}$Mn &  0.25    &  $1^+_1,0$  & 
$|[f_{7/2}],5/2 \rangle \times |[f_{7/2}],5/2 \rangle $    
                                          & 8.2 & 7.2 & 6.7(14)$^2$ \\ 
          &          & $1^+_2,1$  & 
$|[f_{7/2}],5/2 \rangle \times |[f_{7/2}],3/2 \rangle $ 
                                          & 5.6 & 5.0 &   \\ 
          &          & $1^+_3,1$  & 
$|[f_{7/2}],5/2 \rangle \times |[f_{7/2}],7/2 \rangle $ 
                                          & 3.2 & 2.9 &   \\ 

$^{54}$Co &  0.16    &  $1^+_1,0$  & 
$|[f_{7/2}],7/2 \rangle \times |[f_{7/2}],7/2 \rangle $ 
                                          & 14.2 & 12.5 & 12(2)$^2$ \\   
          &           & $1^+_2,1$  & 
$|[f_{7/2}],7/2 \rangle \times |[f_{7/2}],5/2 \rangle $ 
                                          & 3.4 & 3.1 &   \\   
\end{tabular}
\end{center}
\end{table}

\begin{table}
\caption{
Individual B(M1;$0^+_1 \rightarrow 1^+_f,K_f$) and summed 
 $\sum = \sum_f B(M1;0^+_1 \rightarrow 1^+_f,K_f)$ 
values for the three 
lowest $1^+,T=0$ states. Results of calculations using exact formula 
(\ref{m1:general}) and deformed 
single-$j$-orbital approximation formulae  (\ref{bm1:k01}) and 
(\ref{bm1:k11}) are shown. The bare spin $g$-factors were used.
}
\label{tab:sum}
\begin{center}
\begin{tabular}{ccccc}
         State &
\multicolumn{4}{c}{B(M1;$0^+_1 \rightarrow 1^+_f,K_f$), ($\mu_N^2$)} \\ 
      $J^\pi_f,K_f$ &   
\multicolumn{2}{c}{$^{46}$V} & \multicolumn{2}{c}{$^{50}$Mn} \\
\cline{2-3} \cline{4-5}
              & Eq.~(\ref{m1:general}) & 
Eqs.~(\ref{bm1:k01},\ref{bm1:k11}) &  Eq.~(\ref{m1:general}) & 
Eqs.~(\ref{bm1:k01},\ref{bm1:k11}) \\
 \cline{2-3} \cline{4-5}
 $1^+_1,0$  & 3.7  & 2.6  &  8.2 & 7.2   \\    
 $1^+_2,1$  & 7.6  & 8.7  &  5.6 & 6.9  \\
 $1^+_3,1$  & 5.7  & 6.9  &  3.2 & 4.1  \\
\hline 
  $\sum$  & 17.0 & 18.2 & 17.0 & 18.2 
\end{tabular}
\end{center}
\end{table}


\begin{references} 

\bibitem{Vogel} P.~Vogel, Nucl. Phys. A 662 (2000) 148.

\bibitem{Wyss1} W.~Satula, R.~Wyss, Nucl. Phys. A 676 (2000) 120.

\bibitem{Mac2} A.~O.~Macchiavelli {\it et al.}, Phys. Lett. B 480 (2000) 1.

\bibitem{Faes99} A.~Petrovici, K.~W.~Schmid, A.~Faessler, Nucl. Phys. A 647 
                                                 (1999) 197.

\bibitem{Ots98} T.~Otsuka, M.~Honma, and T.~Mizusaki, Phys. Rev. Lett.
                81 (1998) 1588.

\bibitem{Schm} T.~Schmidt, Z. Phys. 106, (1937) 358.

\bibitem{Lis99} A.F. Lisetskiy {\it et al.}, Phys. Rev. C 60 (1999) 064310.

\bibitem{Davidson} R.~J.~Ascuitto, D.A.Bell and J.P.Davidson,  Phys.Rev. 176
                  (1968) 1323. 
\bibitem{Malik} P.~Wasilewski and F.~B.~Malik, Nucl. Phys. A, 160 (1971) 113. 

\bibitem{Bohr} A.~Bohr and B.~R.~Mottelson, {\it Nuclear Structure}, vol.II, 
               Benjamin: New York, 1969.


\bibitem{Guise98} Hubert de Guise, David J.~Rowe, Nucl. Phys. A 636 (1998) 
                  47.  

\bibitem{Irvine} J.~M.~Irvine, {\it Nuclear Structure Theory}, 
                 Pergamon Press: Oxford, 1972.

\bibitem{Bruss77} P.~J.~Brussaard and P.~W.~M.~Glaudemans, {\it Shell-model 
applications in nuclear spectroscopy}, North-Holland Publishing Company, 
               Amsterdam, 1977. 

\bibitem{Fries98} C. Frie\ss{}ner {\it et al.},  Phys. Rev. C 60 (1999) 
                  011304.

\bibitem{Brent01} P.~von Brentano,{\it et al.}, Nucl. Phys. A 682 (2001) 48c.

\bibitem{Pietr00} N.Pietralla {\it et al.}, in preparation.

\bibitem{Schneider99} I. Schneider {\it et al.}, Phys. Rev. C 61 (2000) 
                                044312. 
\bibitem{Schmidt} A. Schmidt {\it et al.}, Phys. Rev. C 62 (2000) 
                                044319.

\bibitem{Zamick2} L.~Zamick, N.~Auerbach, Nucl. Phys. A 658 (1999) 285.


\bibitem{Erice2}  P.~von~Brentano, {\it et al.}, Progr. Part. Nucl. Phys.
 {\bf 46} (2001) (in press).

\bibitem{Wi} M.~Carchidi and B.~H.~Wildenthal, Phys. Rev. C 37 (1988) 1681. 

\bibitem{Bologna} P.~von Brentano,{\it et al.}, 
Proceedings of the "Bologna 2000 - Structure of the Nucleus at the Dawn of 
the Century" Conference; nucl-th/0009043.

\bibitem{NS} P.~M.~Endt, At. Data Nucl. Data Tables 55 (1993) 171. 


\end{references}
\end{document}